\documentclass[twocolumn,showpacs]{revtex4}
\usepackage{graphicx}
\usepackage{dcolumn}
\usepackage{bm}
\usepackage{amsmath}
\usepackage{amsfonts}
\usepackage{amssymb}

\providecommand{\U}[1]{\protect\rule{.1in}{.1in}}

\begin{document}

\title{Air expansion in the water rocket}
\author{Alejandro Romanelli}
\altaffiliation{alejo@fing.edu.uy}
\author{Italo Bove}
\author{Federico Gonz\'alez Madina}
\affiliation{Instituto de F\'{\i}sica, Facultad de Ingenier\'{\i}a\\
Universidad de la Rep\'ublica\\
C.C. 30, C.P. 11300, Montevideo, Uruguay}
\date{\today }

\begin{abstract}
We study the thermodynamics of the water rocket in the thrust phase, taking into account the expansion of the air with
water vapor, vapor condensation and the energy taken from the environment. We set up a simple experimental device with a
stationary bottle and verified that the gas expansion in the bottle is well approximated by a polytropic process
$PV^\beta$= constant, where the parameter $\beta$ depends on the initial conditions. We find an analytical expression for
$\beta $ that only depends on the thermodynamic initial conditions and is in good agreement with the experimental
results.
\end{abstract}

\maketitle

\section{Introduction}

The water rocket is a popular toy that is already several decades old \cite%
{Johnson}. Its inexpensive version can be built with a plastic bottle filled
with water and pressurized air obtained with a bicycle pump. When the bottle
is opened, the internal air pressure pushes the water out and this causes an
increase in the momentum of the bottle. The water ejection gives the bottle
the thrust that allows it to move for several meters, see Ref. \cite{Kagan}.

The launch of water rockets is used in our undergraduate laboratory to illustrate several physics concepts, such as : (a)
Newton laws, (b) conservation of momentum, (c) work-energy theorem, (d) Bernoulli equation, (e) mass conservation
equation and (f) ideal gas expansion. On this last point it is necessary to clarify whether the expansion of the gas in
the rocket is an adiabatic or an isothermal process.

The theoretical description of the water rocket is not trivial and in
general it is necessary to adopt certain premises to obtain a friendly
caricature of reality. In particular, the rocket's only source of energy,
the air expansion, is in general modeled \cite{Kagan,Prusa,Perotti,Perottib}
as an isothermal or an adiabatic process involving dry air. The qualitative
predictions of these theoretical approaches give reasonable results, however
their quantitative description are incorrect, showing that some hypothesis
of the model is inadequate. In particular observing the launch of the water
rocket, we see that when the ejection of the water concludes, an additional
ejection of fog follows. Therefore, it is clear that the expansion occurs
for a mixture of dry air, water vapor and condensed water (fog). In
reference \cite{Gommes} the author deepens in this point and he builds a
pressure-volume relation assuming that the total entropy of the mixture is
conserved during the expansion; he additionally shows that the solution of
the pressure-volume equation can be approximated by a polytropic process.

We developed an experimental device in order to elucidate the nature of the
water-rocket gas expansion. We also present, a brief theoretical development
that provides an analytical expression for the parameter that characterize
the polytropic processes.

In the following we shall use the term ``gas" to refer to the mixture of:
dry air and water vapor. The paper is organized as follows, in the next
section we study the exchange of heat between the gas and the environment in
a polytropic process. The experimental device and our results are presented
in the third section. In the last section some conclusions are drafted.

\section{Polytropic process and exchange of energy}

We present here a brief revision of the polytropic process and the calculation of some quantities of heat involved. A
polytropic process is a quasistatic process carried out in such a way that the specific heat remains constant during the
entire process \cite{definición1,definición2,definición3}. Therefore the heat exchange when the temperature changes is
\begin{equation}
\frac{dQ}{dT}=N c,  \label{dqdt}
\end{equation}
where $Q$ is the heat absorbed by the system, $T$ is the temperature, $c$ is a constant specific heat and $N$ is the
number of moles. Pressure and volume can change arbitrarily in this process, therefore the specific heat also depends of
the process. Of course, the processes in which the pressure or the volume are kept constant are polytropic with specific
heats $c_p$ or $c_v $ respectively. In an adiabatic evolution there is no heat exchange between the system and the
environment, therefore it is a polytropic process with $c=0$. At the other extreme, the isothermal evolution is
characterized by a constant temperature therefore it may be thought as a polytropic process with infinite specific heat.

Using the first law of thermodynamics together with the internal energy of an ideal gas $U=c_v NT$ and Eq.(\ref{dqdt}),
we have for an infinitesimal polytropic process
\begin{equation}
(c_v-c)N dT+P dV=0,  \label{poly}
\end{equation}
where $P$ is the pressure and $V$ the volume. From Eq.(\ref{poly}) and using
the ideal gas equation of state, it is easy to prove that the polytropic
process satisfies
\begin{equation}
P {V}^\beta=constant,  \label{poly2}
\end{equation}
where
\begin{equation}
\beta=\frac{c_p-c}{c_v-c}.  \label{poly3}
\end{equation}
Usually thermodynamic textbooks define a polytropic process through Eq.(\ref{poly2}) (for example \cite{Turns}). However
we have started with Eq.(\ref{dqdt}) which gives a direct link between heat and temperature changes. In particular for an
adiabatic process $c=0$ and $\beta=\gamma\equiv c_p/c_v$ and for an isothermal process, $c=\infty$ and $\beta=1$. For the
process treated in this paper we find that $1<\beta<\gamma$ where $\gamma=1.4$ for dry air at room temperature. The
specific heat as a function of $\beta$ is
\begin{equation}
c=\frac{c_p-\beta c_v}{1-\beta}.  \label{poly4}
\end{equation}
The total heat $Q$ exchanged between the gas and the environment can be
calculated integrating Eq.(\ref{dqdt})
\begin{equation}
Q=c N (T-T_0),  \label{q1}
\end{equation}
where $T$ is the gas temperature and $T_0$ is the environment temperature, which is also the water temperature and the
initial temperature of the gas in the bottle. An experimental measurement of the time dependence of $T$ determines the
time dependence of $Q$ through Eq.(\ref{q1}).

As the water is expelled from the bottle the gas expands and cools. In this
process the relative humidity increases up to saturation and some of the
vapor condenses. The quantity of condensed water vapor at the end of the
experiment will depend on the initial relative humidity in the gas and the
initial and final temperatures. We want to estimate the mass and the heat
involved in this condensation. The mass of water vapor $m_v$ can be
calculated using \cite{Sonntag}
\begin{equation}
m_{v}=\frac{P_{v} V}{R_v T},  \label{vapor}
\end{equation}
where $P_{v}$ is the partial pressure of the vapor at $T$ and $R_v=0.4615$ {kJ/(kg K)} is the ideal gas constant for the
vapor. The vapor partial pressure is determined by de equation $P_v=\varepsilon e_w$ where $\varepsilon\in[0,1]$ is the
fraction of relative humidity and $e_w$ is the saturation pressure which depends only on the temperature and may be
obtained from a vapor table. The pressurized air is generated by the compressor at a higher temperature than the
environment and its cooling results in the saturation of the air with vapor inside the system. Thus we assume that we are
working with $e_w=1$ during all the course of the experiments, then $P_v=e_w$. This assumption implies the maximization
of the energy delivered by the vapor-liquid phase transition. In our case the initial temperature is $T_0=292$ K and
partial pressure is $P_{v_{0}}=2.3$ kPa (this last value was taken from a vapor table).

In order to get a fuller understanding about the heat exchange due to vapor
condensation we start with the Clapeyron-Clausius equation \cite{Reif} for
the vapor-liquid phase transition
\begin{equation}
\frac{d e_w}{d T}={\frac{L_v}{R_v}\frac{e_w}{T^{2}}},  \label{clau0}
\end{equation}
where $L_v$ is the latent heat, which for our purposes may be considered
constant, $L_v=2500$ kJ/kg. The solution of Eq.(\ref{clau0}) gives the
partial pressure $P_v=\varepsilon e_w$ as a function of the temperature $T$
\begin{equation}
P_v=P_{v_{0}} e^{\frac{L_v}{R_v}(\frac{1}{T_0}-\frac{1}{T})},  \label{clau}
\end{equation}
where the constant was adjusted so that $P_v(T_0)=P_{v_{0}}$.

The condensation produces a change in the mass of vapor in the gas. The mass of condensed vapor can be calculated using
Eq.(\ref{vapor}) twice, that is
\begin{equation}
M=\frac{P_{v_{0}}V_{0}}{R_{v}T_{0}}-\frac{P_{v}V}{R_{v}T}.  \label{vapor2}
\end{equation}
It is easy to prove that the polytropic process is also characterized by the equation
$PT^{\frac{\beta}{1-\beta}}=constant$. Then,
\begin{equation}
\frac{V}{T}=\frac{NR}{P}=\frac{NR}{P_0}\left(\frac{T_0}{T}\right)^{\frac{\beta}{\beta-1}},  \label{vapor2a}
\end{equation}
where $R$ is the ideal gas constant. To obtain the latent heat of condensation $Q_{v}=ML_{v}$ we substitute
Eqs.(\ref{clau},\ref{vapor2a}) into Eq.(\ref{vapor2}) that is
\begin{equation}
Q_{v}=\frac{NRP_{v_0}L_{v}}{R_{v}P_{0}}\left[ 1-e^{\frac{L_v}{T_0R_v}(1-%
\frac{T_0}{T})}\left(\frac{T_0}{T}\right)^{\frac{\beta}{\beta-1}}\right]. \label{vapor3}
\end{equation}

One may wonder if there are other types of heat exchange between the system and the environment. As will be seen our
experiment lasts for less than one second, then it is very reasonable to assume that the heat exchanged through the walls
is not important, the energy delivered by the vapor-liquid phase transition remains as the sole source of heat for the
system.

It is important to point out that, due to the water-vapor condensation, the number of moles $N$ in the gas is not
constant. However, Eqs.(\ref{q1},\ref{vapor3}) were obtained using the ideal gas law for air and water vapor, which
assume that $N$ and $N_{v}$ (the number of moles of water vapor) are constants. Therefore, Eqs.(\ref{q1},\ref{vapor3})
may be used only if the variation of moles is negligible during condensation. The relative number of moles condensed
${\delta N_{v}}/N_{v}$ can be estimated using Eqs.(\ref{vapor},\ref{vapor2})
\begin{equation}
\frac{\delta N_{v}}{N_{v}}=\frac{M}{m_{v_{0}}}=1-\left( \frac{T_{0}}{T}%
\right) ^{\frac{\beta }{\beta -1}}~e^{\frac{L_{v}}{R_{v}}(%
\frac{1}{T_{0}}-\frac{1}{T})},  \label{ene}
\end{equation}%
where $m_{v_{0}}$ is the initial water vapor mass. If we estimated the final temperature as $T=T_{0}-10$ K then
Eq.(\ref{ene}) gives us the value ${\delta N_{v}}/N_{v}\sim 0.6$. The total relative variation of moles in the gas
${\delta N_v}/N$ is
\begin{equation}
\frac{\delta N_{v}}{N}=\frac{\delta N_{v}}{N_{v}}\frac{%
N_{v}}{N}=\frac{P_{v_{0}}}{P_{0}}\frac{\delta N_{v}}{N_{v}}\sim 10^{-3}, \label{ene2}
\end{equation}
where we have used the ideal equation gas both for the gas as for the water vapor. Therefore we have the following
situation, ${\delta N_v}/{N}$ is negligible but ${\delta N_{v}}/N_{v}$ is not. In the deduction of Eq.(\ref{vapor3}), we
have used Eq.(\ref{vapor}) which is correct only if ${\delta N_{v}}/N_{v}\sim0$. Then, it is clear that Eq.(\ref{vapor3})
can only be correct in the limit $T\rightarrow T_{0}$.  Therefore, we must substitute Eq.(\ref {vapor3}) by the following
exact differential equation
\begin{equation}
dQ_{v}=\frac{L_{v}}{T_{0}R_{v}}\frac{P_{v_{0}}NR}{P_{0}}\left( \frac{\beta }{%
1-\beta }+\frac{L_{v}}{R_{v}T_{0}}\right) dT,  \label{vapor32}
\end{equation}%
that it is obtained replacing $T=T_{0}+dT$ in Eq.(\ref{vapor3}) and taking the terms to first order in $dT$. Now it is
possible to equate $dQ_{v}$ with $dQ$ taken from  Eq.(\ref{dqdt})
\begin{equation}
dQ=dQ_{v},  \label{qu}
\end{equation}
and using Eqs.(\ref{poly4},\ref{vapor32}) in Eq.(\ref{qu}) lead to
\begin{equation}
\frac{c_{p}/R-\beta c_{v}/R}{1-\beta }=\frac{L_{v}}{T_{0}R_{v}}\frac{P_{v_{0}}}{%
P_{0}}\left( \frac{\beta }{1-\beta }+\frac{L_{v}}{R_{v}T_{0}}\right) .
\label{nueva}
\end{equation}%
This last equation gives a theoretical expression to obtain the constant parameter $\beta $,
\begin{equation}
\beta =\frac{c_{p}/R+\left( L_{v}/R_{v}T_{0}\right) ^{2}(P_{v_{0}}/P_{0})}{%
c_{v}/R+\left( L_{v}/R_{v}T_{0}\right) ^{2}(P_{v_{0}}/P_{0})-\left( L_{v}/R_{v}T_{0}\right) (P_{v_{0}}/P_{0})}.
\label{beta}
\end{equation}%
This is the main theoretical result in this paper and from its analysis we conclude: (a) If $P_{v_{0}}\sim 0$ or
$L_{v}\sim 0$ then $\beta \sim \gamma =\frac{c_{p}}{c_{v}}$, this means that the process is adiabatic for dry air. (b)
$\beta $ depends only on the thermodynamical initial conditions of the gas, this is unexpected since other parameters
like the water ejection time do not appear. Tab.~\ref{t0} shows the theoretical values of $\beta $ Eq.(\ref{beta}), which
we call $\beta ^{\ast }$, computed for our experimental initial conditions.

\begin{figure}[th]
\begin{center}
\includegraphics[scale=0.4, angle=0]{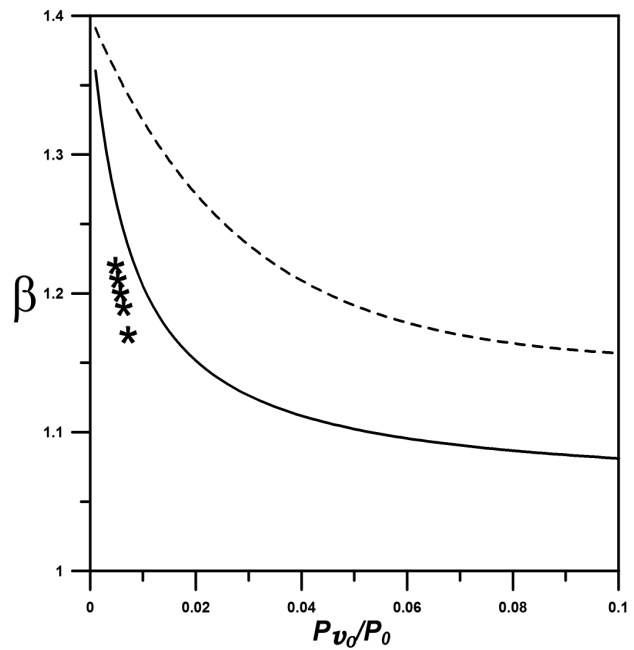}
\end{center}
\caption{The polytropic exponent as a function of the initial molar fraction. The star points are the experimental values
of $\protect\beta $. The solid line is our theoretical calculation given by Eq.(\protect\ref{beta}) and in dashed line is
the function given by Eq.(\protect\ref {betagommes}).} \label{f0}
\end{figure}

It is also interesting to point out that the model developed in Ref.\cite{Gommes} also predicts that the exponent $\beta
$ depends mostly on the initial relative humidity and proposes the empirical function
\begin{equation}
\beta =1.15+(1.4-1.15)\exp (-36P_{v_{0}}/P_{0}).  \label{betagommes}
\end{equation}
Fig.(\ref{f0}) shows $\beta$ (Eqs.(\ref{beta},\ref{betagommes})) as a function of the initial ratio $P_{v_{0}}/P_{0}$
(that is the molar fraction), the figure also shows the experimental values obtained in the present paper.
\section{Experimental results}

In order to elucidate the nature of the expansion of the air in the water
rocket we designed an experimental device to measure the heat exchange
between the compressed gas of the water rocket and the environment. The
water rocket is fixed to the laboratory frame and then the time dependence
of the pressure and volume of the expanding gas in the bottle is measured.
\begin{figure}[th]
\begin{center}
\includegraphics[scale=0.8, angle=0]{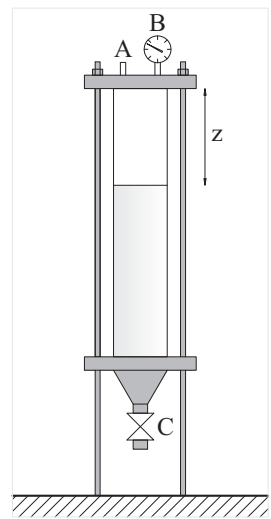}
\end{center}
\caption{The stationary bottle, (A) is a tire valve, (B) is a pressure
manometer, (C) is a faucet, Z is the height of the gas column situated over
the water column.}
\label{f1}
\end{figure}
The stationary bottle is depicted in Fig.~\ref{f1}. The bottle is a transparent acrylic cylinder of inner diameter $5.40$
cm, wall thickness $ 0.30$ cm and length $1$ m, that is maintained in vertical position. The upper and lower covers are
both of aluminum. The upper one has a tire valve and a manometer (WIKA EN $837-1$) working from $0$ to $4$ bar. The lower
cover has a conical shape with a slope of $30$$^{\circ}$ with a $0.5$" faucet that allows to fill and evacuate the water
in the bottle. The bottle is was built in our mechanical workshop. The device is quite simple and the experiment is not
dangerous for the range of pressures used in this paper, so it is appropriate for undergraduate students. In a typical
trial the bottle is filled with about $1400$ cm$^3$ of water and the rest with air. It is placed in vertical position
with the faucet closed. The compressed air is introduced through the valve and then the faucet is opened. The level and
pressure of the gas is filmed with a PIXELINK PL-B$776$F camera. The video was obtained at $50$ frames per second, with a
exposure time of $10 $ ms per frame. The ejection of water from the bottle lasted between $0.64$ s to $0.88 $ s,
depending of the initial pressure, see Tab.~\ref{t0}. Finally the filming was processed manually to obtain the pressure
and volume of the gas as functions of time.

\begin{figure}[th]
\begin{center}
\includegraphics[scale=0.52, angle=0]{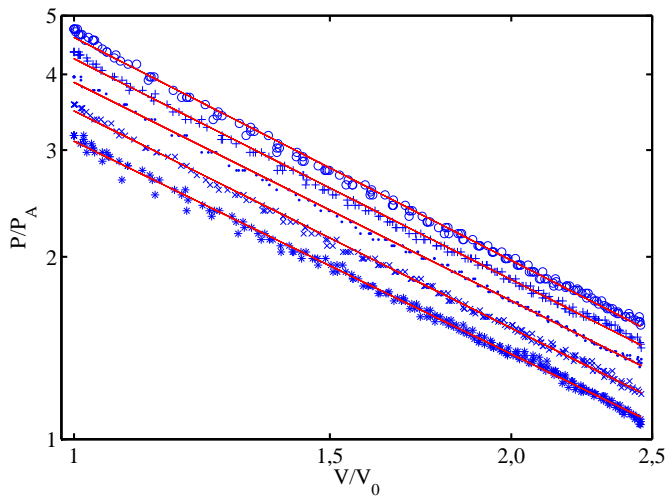}
\end{center}
\caption{The dimensionless pressure as a function of the dimensionless
volume of the gas in a log-log scale. The experimental data for five
experiments are shown together with their adjustment with thin lines. The
values of the initial dimensionless pressure are, from top to bottom, $4.8$,$%
4.4$,$4.0$,$3.6$,$3.2$, with $P_A$ the atmospheric pressure and $V_0=928.6$
cm$^3$.}
\label{f2}
\end{figure}

Fig.~\ref{f2} shows the results for five sets of experimental data of pressure and volume. Each set corresponds to
different initial pressures with the same initial volume; for each initial pressure the experiment was repeated five
times and all the data are incorporated in the same figure. The data are adjusted with a power law $PV^\beta=$constant.
The values of $\beta$, the slope of the adjustment in the figure, are given in Tab.~\ref{t0}. The experimental values of
the polytropic exponents $\beta$ are in very good agreement with their theoretical values $\beta^{*}$ given by
Eq.(\ref{beta}).

\begin{table}[th]
\begin{center}
\begin{tabular}{|c|c|c|c|c|c|}
\hline $P_0/P_A$ & $\beta$ & $\beta^{*}$ & $c/R$ & t(s) & $Q$ (J) \\ \hline
4.8 & 1.22 & 1.27 & -2.0 & 0.64 & 161.7 \\
4.4 & 1.21 & 1.26 & -2.2 & 0.66 & 158.3  \\
4.0 & 1.20 & 1.25 & -2.6 & 0.75 & 154.7  \\
3.6 & 1.19 & 1.24 & -2.7 & 0.75 & 144.8  \\
3.2 & 1.17 & 1.23 & -3.5 & 0.88 & 144.9 \\ \hline
\end{tabular}%
\end{center}
\caption{{}}
\label{t0}
\end{table}
The fifth column of Tab.\ref{t0} shows the ejection time at which end the water is exhausted from the bottle. The heat
$Q$ at the end of the process shown in the last column of Tab.\ref{t0}, it is calculated using Eq.(\ref{q1}) the
experimental data and the ideal gas equation.

\section{Discussion and conclusion}

The typical ejection time of the water in our experiments is $0.7$ s and the
typical ejection time in the usual water rocket is $0.1 $ s \cite{Kagan}.
The difference in the ejection rates must be analyzed: (\emph{i}) The main
geometrical difference between the bottles is related to the nozzles. The
nozzle section of the stationary bottle $a_{1}$ and the standard soda bottle
$a_{2}=3.46~10^{-4}$ m$^2$ are linked by $a_{1} =0.4 a_{2}$. A smaller
nozzle was necessary in order to accommodate the capability of our video
camera. (\emph{ii}) Another difference between the experiments is the
apparent gravity. In the stationary bottle the water is only subjected to
the weight. In the water rocket, the water mass is subjected to a force
which is the sum of its weight plus a fictitious force due to the rocket
acceleration. The apparent gravitational field is several times larger than
the Earth's gravitational field. (\emph{iii}) Finally, our theoretical
calculation shows that $\beta^{*}$ does not depend on the water output rate,
it only depends on the thermodynamic initial conditions of the gas.
Additionally $\beta^{*}$ agrees with our experimental values. Then we
conclude that same process should occur in the water rocket with the same
thermodynamic initial conditions.

In summary, we have presented an experimental study of the air expansion in the water rocket for the first time, using a
stationary bottle. It is shown that the air expansion follows a polytropic process of the type $PV^\beta=constant$ where
$\beta$ is an average value with slight dependence on the initial conditions. Furthermore we obtained an analytical
expression for the polytropic exponent which agrees reasonably with the experimental values. This theoretical exponent
only depends on the thermodynamic initial conditions of the gas and it is also suitable to study the air expansion in the
usual water rocket. Finally we conclude that the vapor condensation is the main energy source for the gas expansion
process in the water rocket.

A.R. and I.B. acknowledge stimulating discussions with V\'{\i}ctor
Micenmacher, Gast\'{o}n Ayub\'{\i} and Pedro Curto, and the support of
PEDECIBA and ANII. F.G.M. acknowledges the support of PEDECIBA through a
scientific initiation scholarship.

\end{document}